\documentclass[twocolumn,a4paper,showpacs]{revtex4}
\usepackage{graphicx}
\usepackage[all]{xy}
\usepackage{amsmath}
\usepackage{color}
\usepackage{amssymb}
\usepackage{tensor}
\newcommand{\be}{\begin{equation}}
\newcommand{\ee}{\end{equation}}
\newcommand{\ben}{\begin{eqnarray}}
\newcommand{\een}{\end{eqnarray}}
\newcommand{\bes}{\begin{subequations}}
\newcommand{\ees}{\end{subequations}}
\def\bal#1\eal{\begin{align}#1\end{align}}

\newcommand{\sech}{{\rm sech}}

\begin{document}

\title{Novel connection between lump-like structures and quantum mechanics}

\author{D. Bazeia}\affiliation{Departamento de F\'\i sica, Universidade Federal da Para\'\i ba, 58051-970 Jo\~ao Pessoa, PB, Brazil}
\author{L. Losano}\affiliation{Departamento de F\'\i sica, Universidade Federal da Para\'\i ba, 58051-970 Jo\~ao Pessoa, PB, Brazil}
\author{Gonzalo J. Olmo}\affiliation{Departamento de F\'\i sica Te\'orica and IFIC, Centro Mixto Universidad de
Valencia - CSIC\\ Universidad de Valencia, Burjassot-46100, Valencia, Spain}


\begin{abstract}{This work deals with lump-like structures in models described by a single real scalar field in two-dimensional spacetime. We start with a model that supports lump-like configurations and use the deformation procedure to construct scalar field theories that support both lumps and kinks, with the corresponding stability investigation giving rise to new physical systems. Very interestingly, we find models that support stable topological solutions, with the stability potential being able to support a tower of non-negative bound states, generating distinct families of potentials of current interest to quantum mechanics. We also describe models where the lump-like solutions give rise to stability potentials that have the shape of a double-well.}
\end{abstract}


\maketitle

\section{Introduction}

The study of defect structures in models governed by real scalar fields in $(1,1)$ spacetime dimensions has been implemented in a diversity of contexts in high energy physics \cite{B1,B2}. These structures can have topological or nontopological behavior and are in general called kinks or lumps, respectively; see, e.g., Refs.~\cite{B1,B2,Bor,Vacha,ave1,ave2,kha} and references therein. An interesting issue that follows from the investigation of the stability of kinks and lumps is that the linear fluctuations around them lead to the presence of potentials that may be used to describe problems in quantum mechanics \cite{IH,Khare,B3,B4,BB1,BB2,BL}.

The connection of kinks and lumps with quantum mechanics (QM) is a well-known fact which has been successfully exploited to describe several phenomena in nonlinear science. In particular, the lump-like structures which we will be studying in the current work may contribute to the formation of structures in the early Universe \cite{B1,stru1,stru2} and q-balls \cite{qb1,qb2,qb3,qb4,qb5}, vortons \cite{v1,v2} and oscillons \cite{O1,O2,O3} in high energy physics. They can also be used to describe bright solitons in applications in optical fibers \cite{OS1,OS2,OS3}, tunneling effects in quantum mechanics \cite{book}, and to model tachyon condensation and lump solutions in string theory \cite{sen1,sen2,MZ,Jak,st1,st2,st3}. Since the lump-like structures are in general unstable, one can also use them to model dissipation, making contact with two interesting studies concerning the presence of dissipation in quantum gravity \cite{gra} and in classical mechanics \cite{cm}, among other interesting possibilities.

The case of q-balls explored in \cite{qb1,qb2,qb3,qb4,qb5}, for instance, concerns models described by charged scalar fields, with the charge being crucial to make the localized solution stable against decay into the elementary excitations present in the models. In this sense, the lump-like structures to be considered in this work may be stabilized by making the scalar field complex, or yet, by coupling other charged fields to them, bosonic or fermionic. These possibilities may add charge to the lump-like structures, and may make them stable, enlarging the scope of the current work.

The purpose of this paper is to use the deformation procedure described in \cite{BLM} to extend the analysis initiated in \cite{BB1,BB2,BL} to explore new scenarios involving lumps and their relation with QM systems. 
The families of models considered in this work include cases that support both lump-like and kink-like configurations, potentially leading to novel applications in the various scenarios mentioned above. The QM counterpart of the field theory models considered here may generate towers of non-negative bound states and potentials with the shape of a double-well. 

We remark that most of the systems to be studied below are unstable against large field fluctuations, so the results may not have direct interest to problems in quantum field theory. However, the classical investigations which we introduce are of interest to QM and more, in the context of effective field theory the models may find applications to describe features of localized structures like q-balls, vortons, oscillons and tachyon condensation, as we have already commented on above. Yet in the context of effective field theory, the current work explore classical issues that are similar to ideas explored before in several works, in particular in \cite{Z}, in which the lump solution of $\phi^3$ field theory provides a toy model for unstable D-branes of bosonic string theory, in \cite{V1,V2}, which deal with localized structure in a scalar field model described by a vacuumless potential, in Refs.~\cite{Ba1,Ba2}, which investigate the presence of topological solutions in scalar field theory with a non canonical kinetic term, in which the kinematics is modified to include higher power in the derivative of the scalar field, and in Ref.~\cite{bh}, where the non canonical modification of the kinematics is used to explore the formation of compact structures similar to kinks and vortices.

In this paper, we follow similar ideas and explore models that support localized lump-like structures and their deformations into kink-like structures. The content of the paper is organized as follows. In Sec. II we review the inverted $\phi^4$ model and show that it supports lump-like configurations that are unstable against small fluctuations. We then use in Sec. III the methods developed in \cite{BLM} to generate new families of models that support lumps and kinks, with the kink-like configurations inducing the presence of towers of non-negative bound states. In Sec. IV a different model is introduced, which gives rise to QM potentials with the double-well profile. We end the work with some comments and conclusions in Sec. V. 

\section{Lump-like structure}
\label{met}

\begin{figure}[t]
\centerline{\includegraphics[{height=2.8cm,width=7cm}]{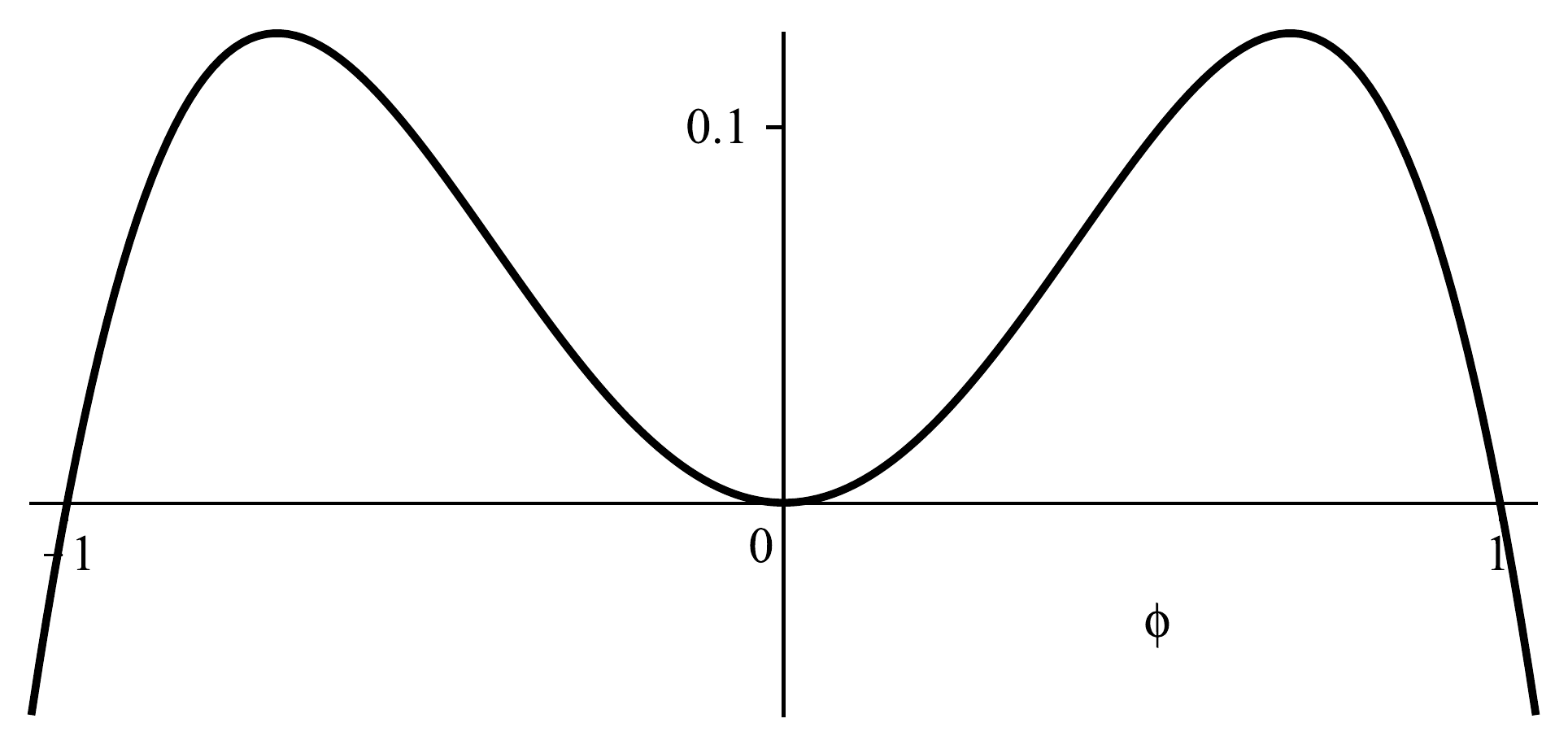}}
\caption{The inverted $\phi^4$ potential \eqref{v1}.}\label{fig1}
\end{figure}

Let us deal with a real scalar field $\phi=\phi(x,t)$ in $(1,1)$ spacetime dimensions, described by the Lagrangian density 
\be\label{model}
{\cal L}=\frac12\partial_\mu\phi\partial^\mu\phi-V(\phi).
\ee
We use natural units $(\hbar=c=1)$ and consider dimensionless field and coordinates; also, we take the metric such that $x^\mu=(x^0=t,x^1=x)$ and focus on the inverted $\phi^4$ model, which is defined by the potential
\be\label{v1}
V(\phi)=\frac12\phi^2(1-\phi^2).
\ee
It has two nontopological sectors, the left and right sectors which are illustrated in Fig.~\ref{fig1}. Although the two sectors are not symmetric around the corresponding maxima, the lump-like solutions are symmetric because they start at the minimum at $\phi=0$ and go to the zero of the potential (at $\phi=\pm 1$) and then return to the minimum at $\phi=0$. The static solutions are bell-shaped, known as lump-like configurations, and here they are given by $\phi_l(x)=-{\rm sech}(x)$ and $\phi_r(x)={\rm sech}(x)$.

\begin{figure}[t]
\centerline{\includegraphics[{height=3cm,width=7cm}]{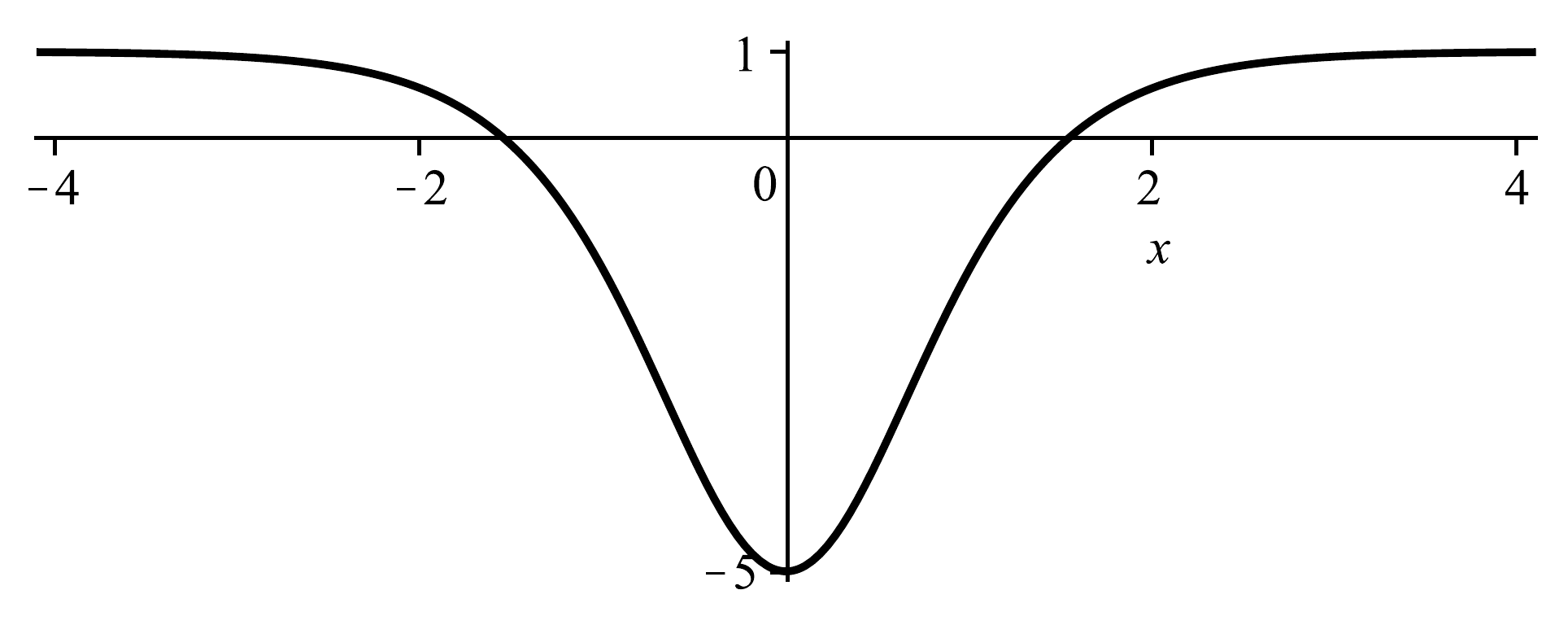}}
\caption{The modified P\"oschl-Teller potential \eqref{u1}.}\label{fig2}
\end{figure}

If one investigates linear stability of these lump-like solutions, 
in the form $\phi(t,x)=\phi(x)+\eta(t,x)$ with
$\eta(t,x)=\sum \cos(w_n\,t)\, \eta_n(x)$, one gets a Schr\"odinger-like equation which can be written as 
\be\label{sch}
\left(-\frac{d^2}{dx^2}+u(x)\right)\eta_n (x)=w_n^2\;\eta_n (x)\,,
\ee
where $u(x)=V_{\phi\phi}$ evaluated on the static field solution $\phi(x)$. In the case under consideration the stability potential $u(x)$ is given by
\be\label{u1}
u(x)=1- 6\,{\rm sech}^2(x).
\ee
It is of the modified P\"oschl-Teller type \cite{MF} and is illustrated in Fig.~\ref{fig2}. It has two bound states, the zero mode that has zero energy and another bound state, with negative energy, which is dimensionless and equals $-3$ for the above model. As it is well-known, the presence of the negative energy bound state makes the lump-like solutions linearly unstable. Anyway, the investigation of stability of the defect structures is directly connected with quantum mechanics, so it is also of interest to quantum physics; see, e.g., Refs.~\cite{Khare,B3,B4,book}. 

In order to describe the topological behavior of lumps and kinks, we introduce the topological current in the standard way \cite{B1,B2}
\be
j^\mu=\varepsilon^{\mu\nu}\partial_\nu\phi.
\ee
It is conserved and induces the topological charge $Q=\phi(\infty)-\phi(-\infty)$, which only depends on the asymptotic behavior of the solution. It is nonzero for kinks; however, it vanishes in the case of lumps, so they are not topologically protected and may decay to the constant and uniform configuration which identifies the minimum of the potential, which has the same vanishing topological charge.

\begin{figure}[t]
\centerline{\includegraphics[{height=3cm,width=7cm}]{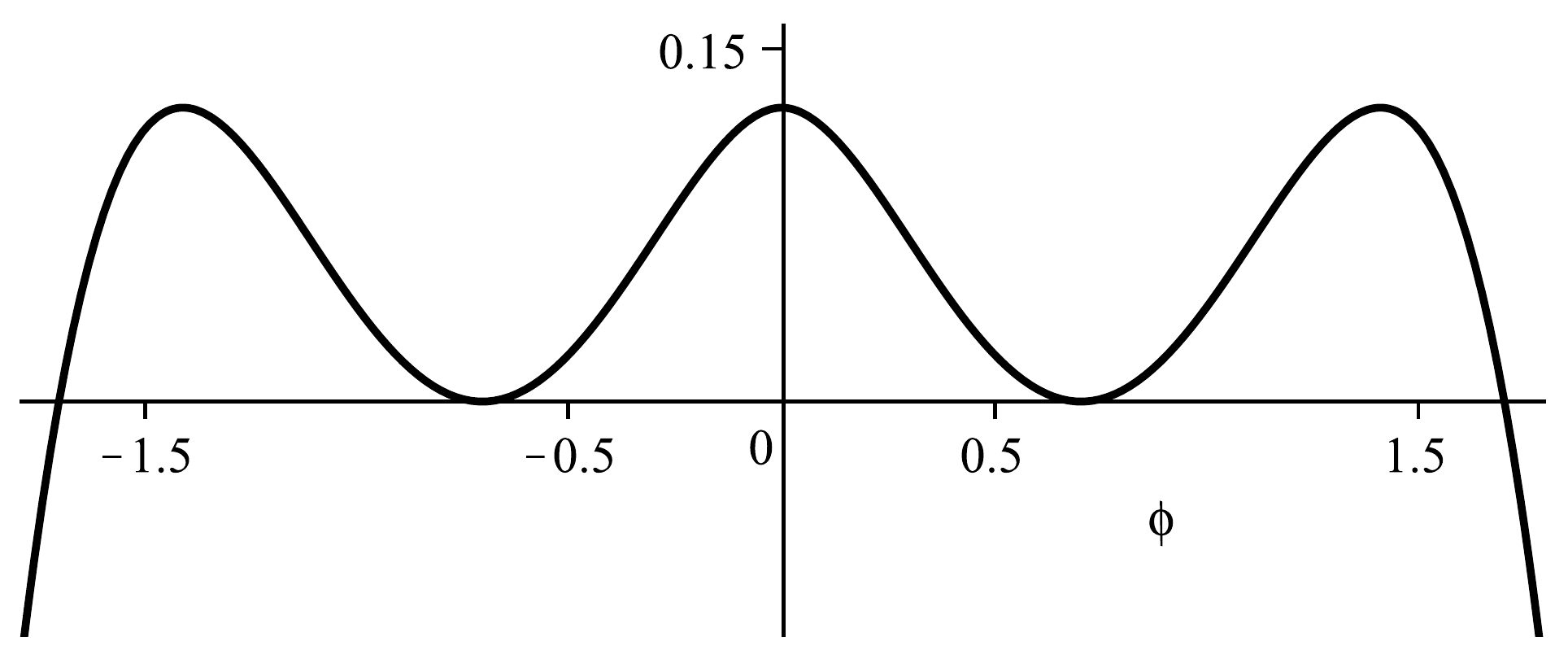}}
\caption{The potential \eqref{v2}, depicted for $a=\sqrt{2}/2$, showing its two lateral nontopological sectors and the new central topological sector.}\label{fig3}
\end{figure}

\section{The procedure}
\label{pro}

The inverted $\phi^4$ model introduced above can be used as the basis to generate other models using the deformation method of \cite{BLM}. Paralleling the analysis of \cite{BL}, we consider several distinct possibilities below. 

\subsection{From lumps to stable intermediate kinks}

We first take the deformation function
\be\label{def1}
f_a(\phi)=a-|\phi|\,,
\ee
with $a$ being real and positive, and use it in the potential \eqref{v1} to get to
\be\label{v2}
V_a=\frac12(a-|\phi|)^2(1-a^2+2a|\phi|-\phi^2)\,.
\ee
The potential \eqref{v1} has two maxima at $\phi=\pm\sqrt{2}/2$. Then, the potential \eqref{v2} is a smooth function for $a=0$ and for $a=\sqrt{2}/2$. For $a=0$ we get back to the potential \eqref{v1}, but for $a=\sqrt{2}/2$ one gets a new potential, which has two lateral nontopological sectors and a central topological sector, as illustrated in Fig.~\ref{fig3}. 

The deformation procedure which we used to write the potential $V_a$ in Eq.~\eqref{v2} was proposed in Ref.~\cite{BLM}. Mathematically, one writes
\be
V_a(\phi)=\frac{V(\phi\to f_a(\phi))} {f^{\prime 2}(\phi)},
\ee
and the recipe is to write the new potential $V_a$ starting with the potential $V(\phi)$ giving by \eqref{v1}, changing $\phi$ by $f_a(\phi)$ and then dividing it by the square of the derivative of the deformation function.

The  lump-like solutions of the potential \eqref{v2} for $a=\sqrt{2}/2$ are $\phi_\pm(x)=\pm\sqrt{2}/2\pm{\rm sech}(x)$, for the two nontopological sectors, the left ($-$) and right ($+$) sectors. Moreover, for the central topological sector we also have the kinklike solutions
\begin{equation}\label{sol2k}
{\phi}(x)=\pm
\left\{
\begin{array}{c}
-\frac{\sqrt{2}}{2}+{\rm sech}(x-x_0),\;\;\;\;\;x\leq 0\,,\\
\,\,\\
\,\,\,\,\frac{\sqrt{2}}{2}-{\rm sech}(x+x_0),\;\;\;\;\;x\geq 0\,,
\end{array}
\right.
\end{equation}
where $x_0={\rm arcsech}({\sqrt{2}/2})$.

\begin{figure}[t]
\centerline{\includegraphics[{height=3.0cm,width=7cm}]{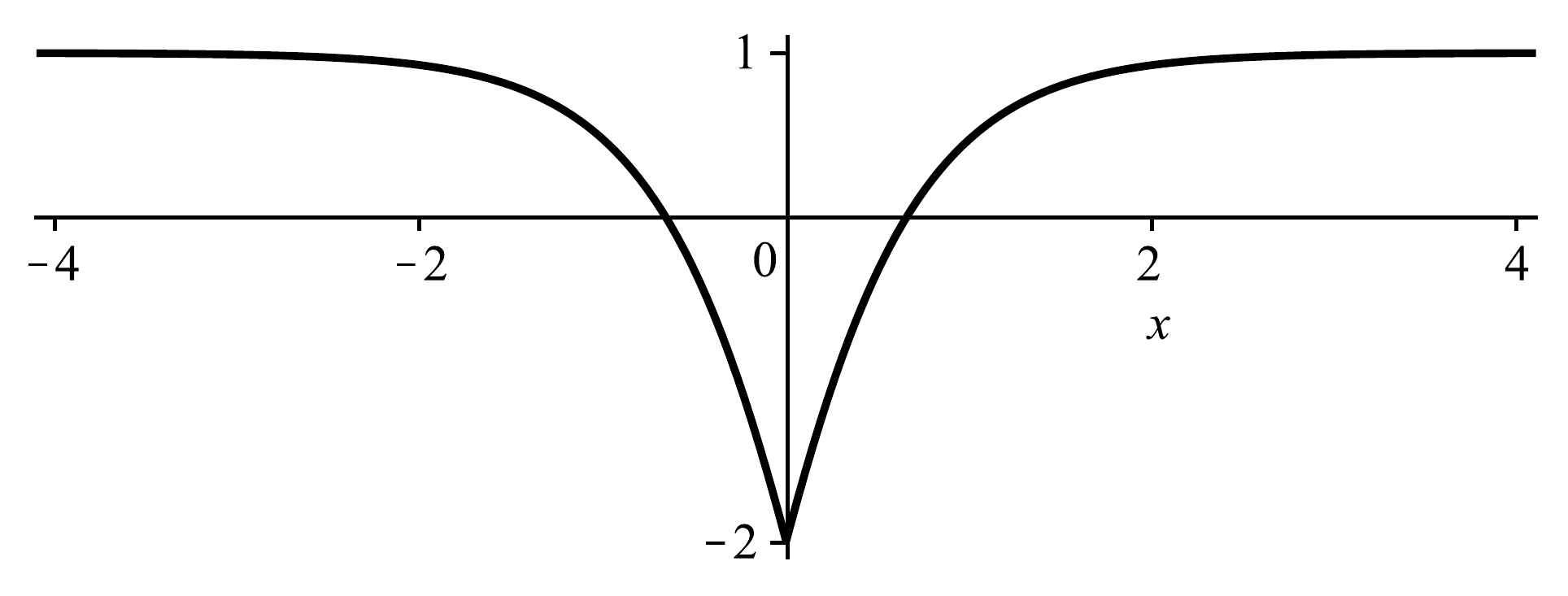}}
\caption{The stability potential \eqref{u2}.}\label{fig4}
\end{figure}
\begin{figure}[h!]
\centerline{\includegraphics[{height=3.0cm,width=6cm}]{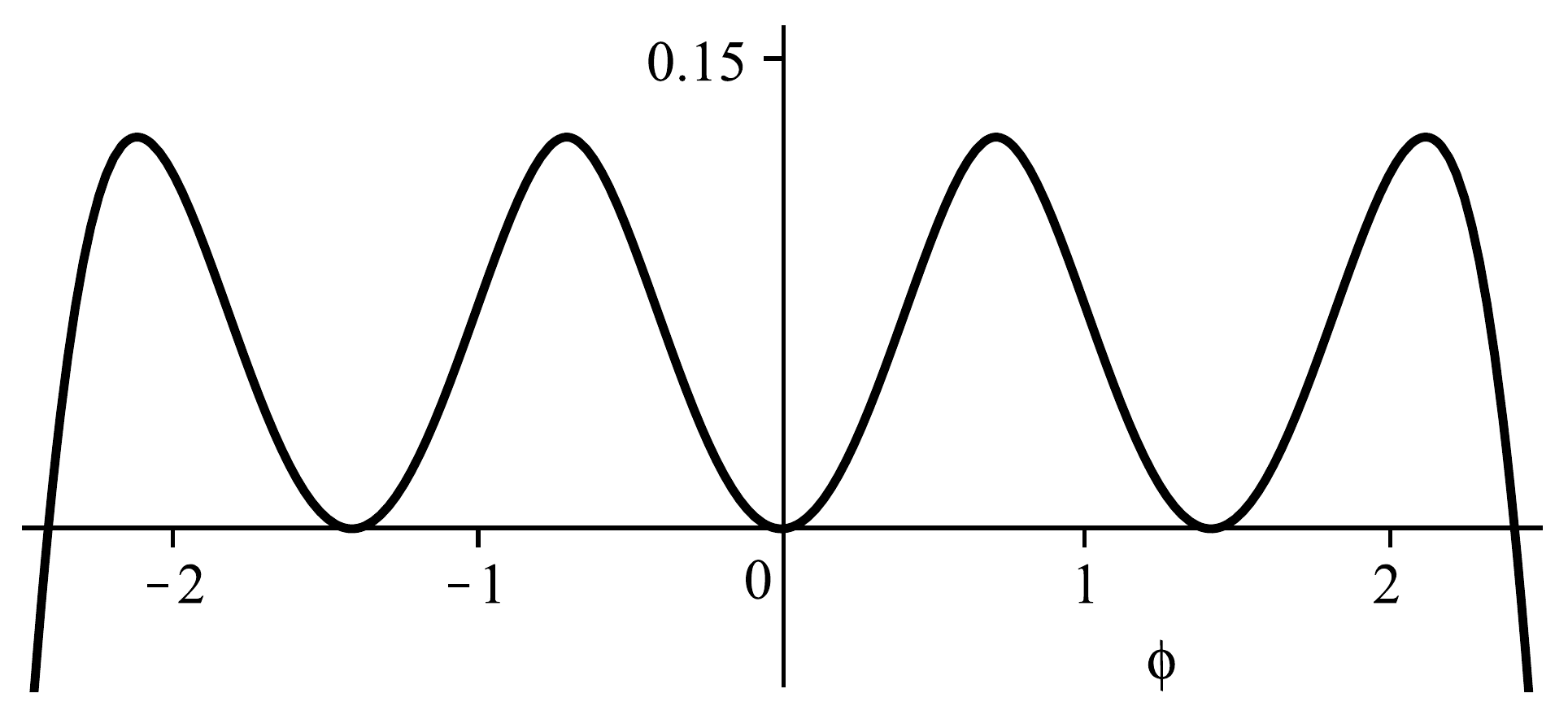}}
\centerline{\includegraphics[{height=3.0cm,width=8cm}]{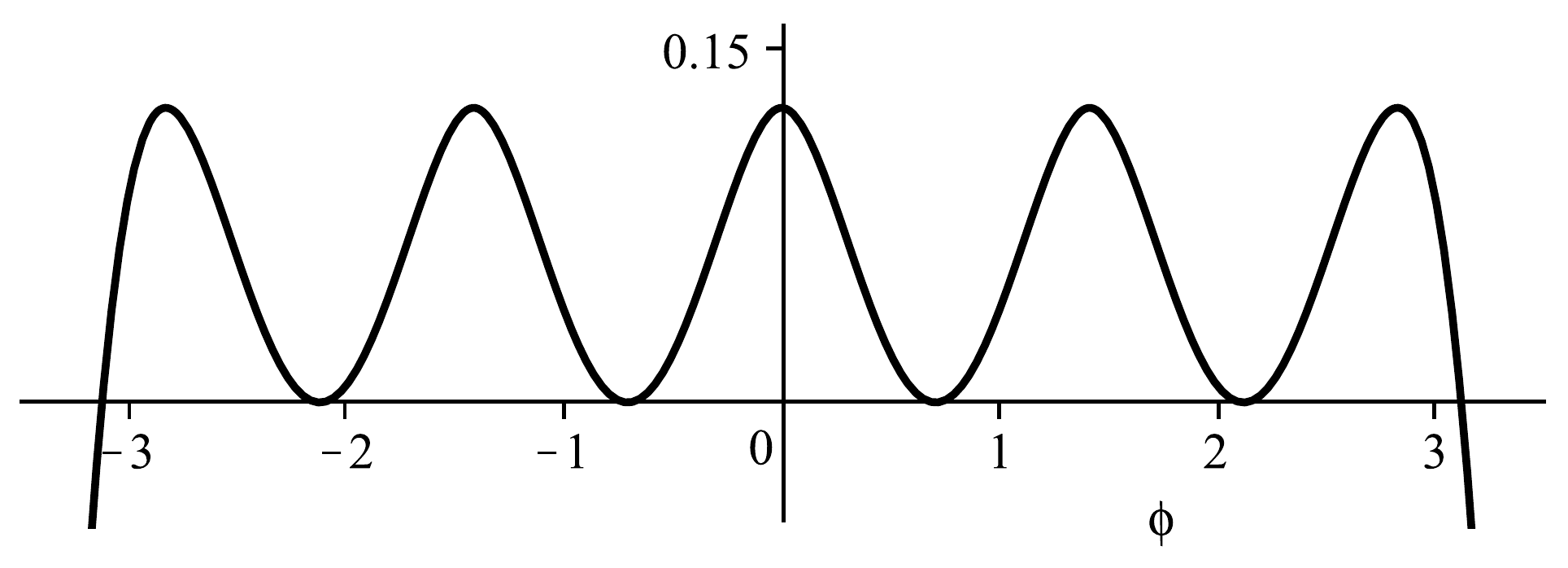}}
\caption{The deformed potentials obtained applying the deformation \eqref{def1} with $a=\sqrt{2}/2$ in the inverted $\phi^4$ potential \eqref{v1} two and three times, in the top and bottom panels, respectively. }\label{fig5}
\end{figure}

If one considers the two lateral nontopological sectors, one obtains the same quantum potential \eqref{u1}, as in the case of $a=0$. However, the central topological sector is new and gives rise to a new stability potential. It has the form
\begin{equation}\label{u2}
u(x)=\pm
\left\{
\begin{array}{c}
1-6\,{\rm sech}^2(x-x_0),\;\;\;\;\;x\leq 0\,,\\
\,\,\\
1-6\,{\rm sech}^2(x+x_0),\;\;\;\;\;x\geq 0\,,
\end{array}
\right.
\end{equation}
as illustrated in Fig.~\ref{fig4}. This potential is shallower than (4) and, as a result, it supports only one bound state, the zero-mode with vanishing energy. The field configuration is, therefore, stable.

Applying the deformation procedure \eqref{def1} for $a=\sqrt{2}/2$ again and again on the potential \eqref{v1}, one generates new potentials, as illustrated in Fig.~\ref{fig5} for two and three repetitions. One notes that for an odd number of repetitions, the new potential gets an odd number of topological sectors in between the two lateral sectors, and if the number of times is even, one gets an even number of topological sectors in between the two lateral sectors. But they all give rise to the same stability potential \eqref{u1} for the lateral nontopological sectors,
and the potential \eqref{u2} for the central topological sectors.

\begin{figure}[h!]
\centerline{\includegraphics[{height=5cm,width=7cm}]{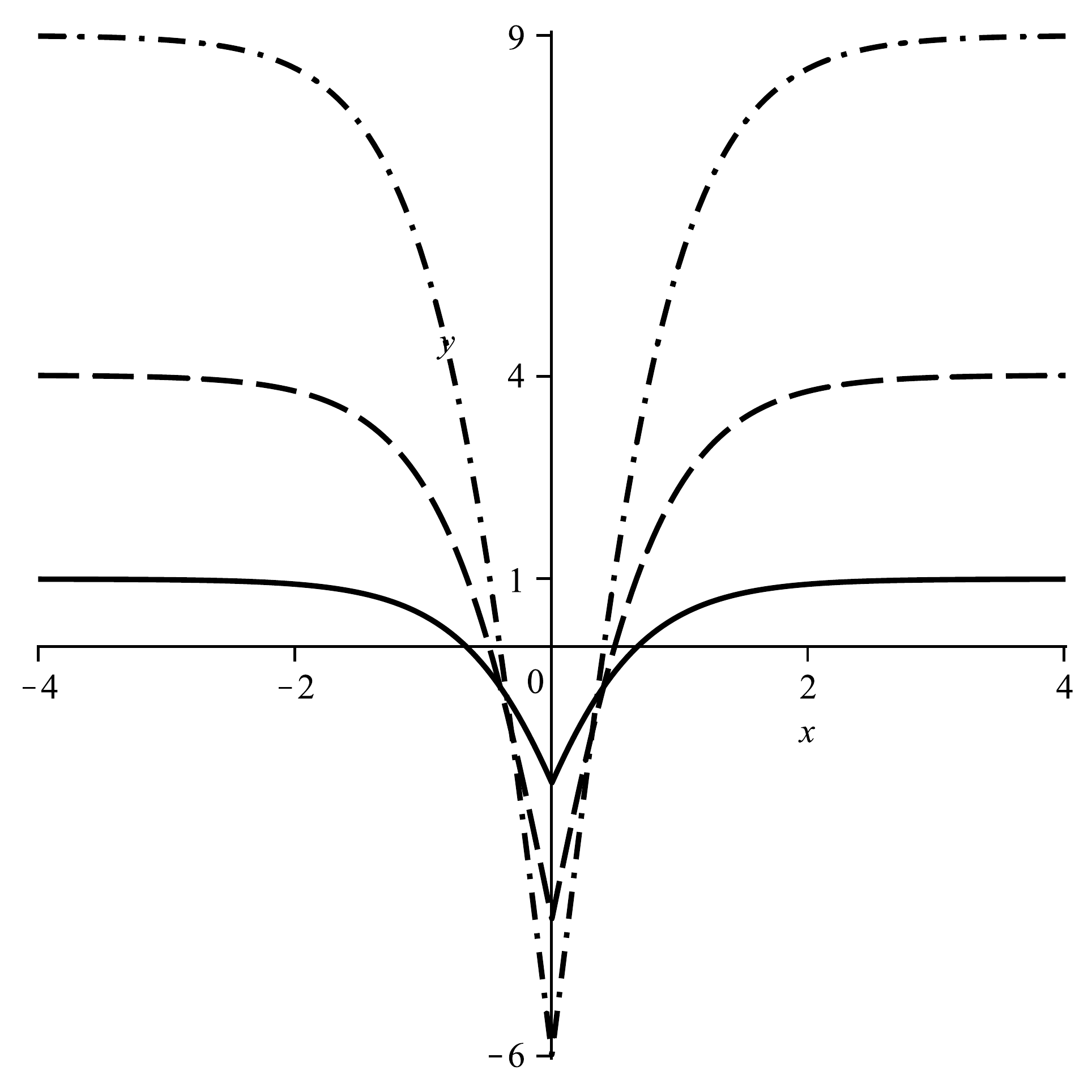}}
\caption{The stability potential \eqref{udn}, depicted for $n=1$ (solid line), $n=2$ (dashed line) and $n=3$ (dotted-dashed line).}\label{fig6}
\end{figure}

\subsection{String-inspired models}

Let us now propose new models, suggested by the study of tachyon condensation in string theory \cite{sen1,sen2}. We consider the inverted $\phi^4$ model given by \eqref{v1} and take advantage of the deformation procedure described in \cite{BLM} to deform it with the deformation function
\be 
f_n(\phi)=|\phi|^{\frac1n},
\ee
with $n=1,2,3,...$\,. This leads us with the family of potentials 
\be\label{vn}
V_n(\phi)=\frac12\,n^2\phi^2\left(1- |\phi|^{\frac2n}\right)\,,
\ee
which have the lump-like solutions $\phi(x)=\pm\,{\rm sech}^{n}(x)$. These potentials have a local minimum at $\phi=0$ and two maxima at $\phi=\pm (n/(n+1))^{{n}/{2}}$, and the same profile as for $n=1$, already illustrated in Fig.~\eqref{fig1}. They have been considered before in \cite{MZ,Jak} to describe tachyon condensation in string theory \cite{sen1,sen2}.

The above models generate the family of stability potentials 
\be\label{un}
u_n(x)=n^2-(n+1)(n+2)\,{\rm sech}^2(x)\,.
\ee
The above modified Poschl-Teller potentials \cite{MF} have $n+1$ bound states, with energy $\epsilon^n_m=n^2-(n+1-m)^2$ for $m=0,1,2,...$, such that
$m\leq n$. We note that: $\epsilon^n_0=-2n-1$ is the negative energy bound state, which depends on $n$; $\epsilon^n_1=0$ represents the zero mode for any $n$; and for $n>1$, there are other $n-1$ bound states with positive energies. The nonnegative energy bound states are as follows: for $n=1$, there is the zero mode; for $n=2$, there are the zero mode and another bound state with energy $3$; for $n=3$, there are the zero mode and two other bound states with energies $5$ and $8$; for $n=4$, there are the zero mode and three other bound states with energies $7$, $12$, and $15$, etc. The spacing between the nonnegative energy levels goes as follows: it is $3$ for $n=2$; $5$ and $3$ for $n=3$; $7$, $5$ and $3$ for $n=4$; etc. One then notices that for a given $n$, the spacing between consecutive energy levels diminishes in a way very different from the case of the energy levels of the hydrogen atom, for instance. 

We follow the previous work \cite{BL} and use the deformation
\be\label{defn}
f_{a_n}(\phi)=a_n-|\phi|\,,
\ee
where $a_n= (n/(n+1))^{{n}/{2}}$, to obtain new family of potentials of the form
\be\label{vdn}
V_{a_n}(\phi)=\frac12n^2\left(a_n-|\phi|\right)^2\left(1- \left|a_n-|\phi|\right|^{\frac2n}\right)\,.
\ee
They have profile similar to the cases already shown in Fig.~\ref{fig3}. The lump-like solutions of the above potential \eqref{vdn} are
$\phi_\pm(x)=\pm\, a_n\pm{\rm sech}^n(x)$, for the two nontopological lateral sectors, and for the central topological sector one gets the kinklike solutions
\begin{equation}\label{solnk}
{\phi}_n(x)=\pm
\left\{
\begin{array}{c}
\!-(n/(n+1))^{{n}/{2}}+{\rm sech}^n(x-x_0),\;\;x\leq 0\,,\\
\,\,\\
\,\,\,\,\!(n/(n+1))^{{n}/{2}}-{\rm sech}^n(x+x_0),\;\;x\geq 0\,,
\end{array}
\right.
\end{equation}
where $x_0={\rm arcsech}(\sqrt{n/(n+1)})$.

We see that for the two lateral nontopological sectors we obtain the same stability potential \eqref{u1}, but for the central topological sector we have 
\begin{equation}\label{udn}
u(x)=\pm
\left\{
\begin{array}{c}
\!n^2-(n+1)(n+2)\,{\rm sech}^2(x-x_0)\,,\;\;x\leq 0\,,\\
\,\,\\
\!n^2-(n+1)(n+2)\,{\rm sech}^2(x+x_0)\,,\;\;x\geq 0\,,
\end{array}
\right.
\end{equation}
as illustrated in Fig.~\ref{fig6}, for $n=1,2,3$. These potentials have bound states that are controlled by $n$, with the zero-mode as the ground-state. The number of bound states is equal to $n$, that is, for $n=3$, for instance, the potential supports the zero mode and two other positive energy bound states, with the same energies engendered by the positive energy bound states of the potential \eqref{un}, as already commented on below Eq.~\eqref{un}. Thus, we have found a topological sector that supports the zero mode and a finite number of bound states, which is a good novelty since this tower of non-negative bound states may find applications in situations of practical interest in supersymmetric QM.

It is interesting to notice that the potential \eqref{udn}, depicted in Fig.~\ref{fig6}, has $n$ bound states, that is, it excludes one bound state, exactly the lowest (negative) energy bound state that appears in the case of lumps. In this sense, the procedure works nicely since  the kink-like solution is stable, so it cannot have negative energy bound state. This is similar to the previous case, since the potential \eqref{u2}, depicted in Fig.~\ref{fig4}, is also shallower and does not support negative bound state.

We can go on and deform the deformed model \eqref{vdn} again and again, generating a family of models with two lateral nontopological sectors, and several topological sectors that are equal to the central topological sector uncovered from the deformation of the model \eqref{vn}.

\section{Another family of models}
\label{sec:new}

\begin{figure}[t]
\begin{center}
{\includegraphics[{height=2cm,width=3cm}]{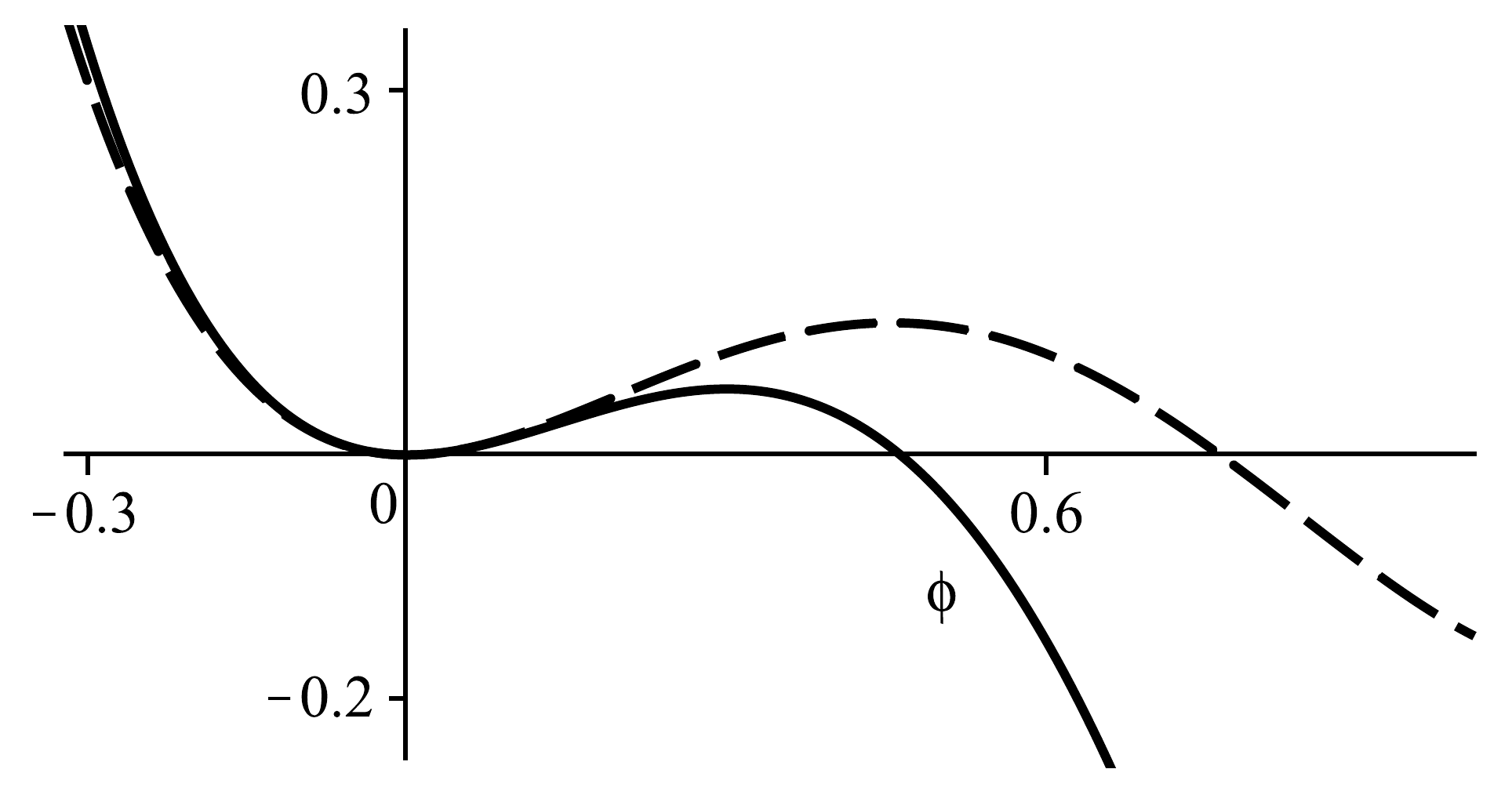}}
\\
{\includegraphics[{height=4cm,width=7cm}]{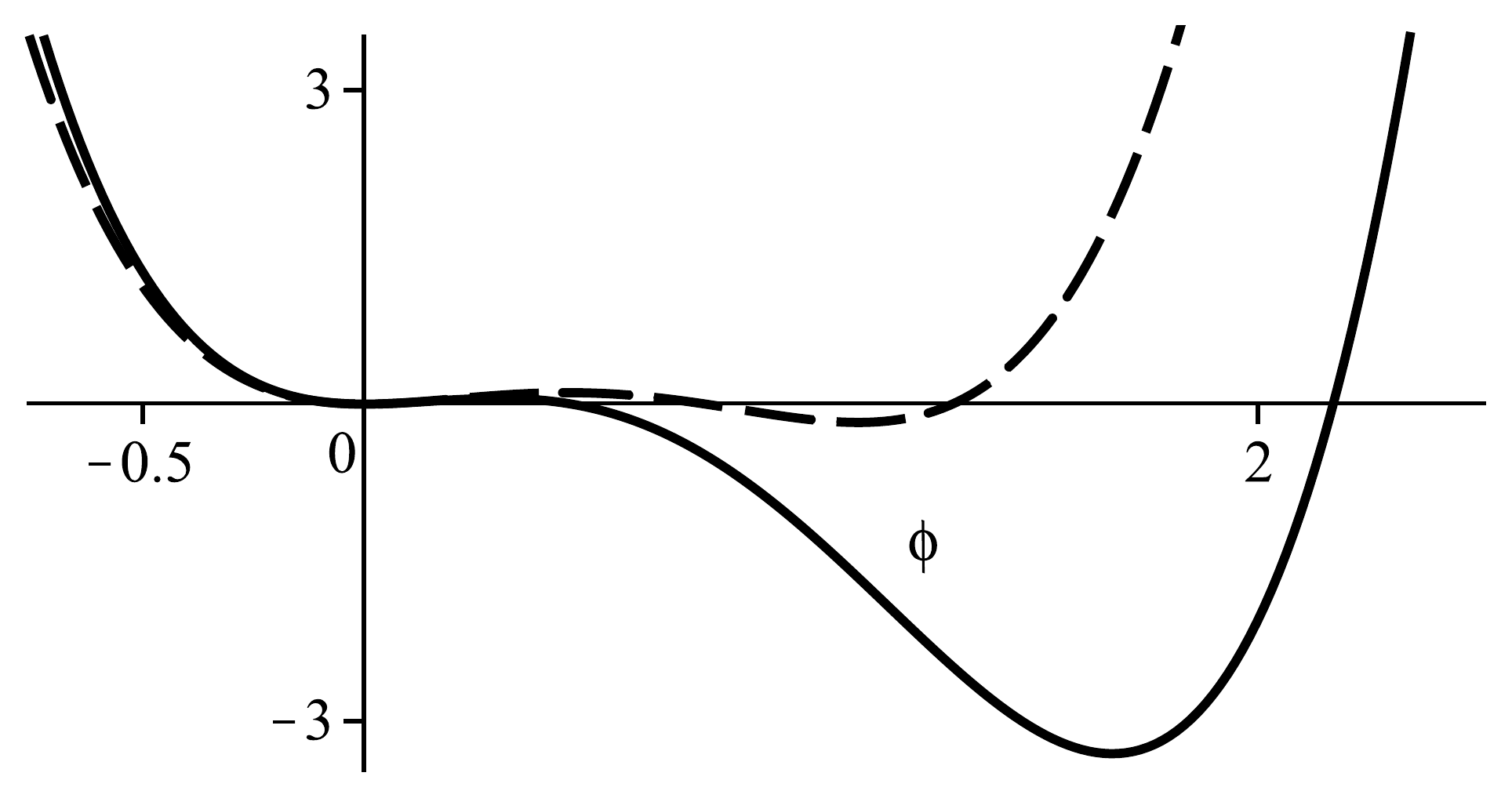}}
\end{center}
\caption{The potential \eqref{vs}, depicted for $s=0.5$ (solid line) and $s=1.0$ (dashed line). The inset at the top panel highlights the behavior of  the potential near the origin.}\label{fig7}
\end{figure}
\begin{figure}[h!]
\centerline{\includegraphics[{height=3cm,width=7cm}]{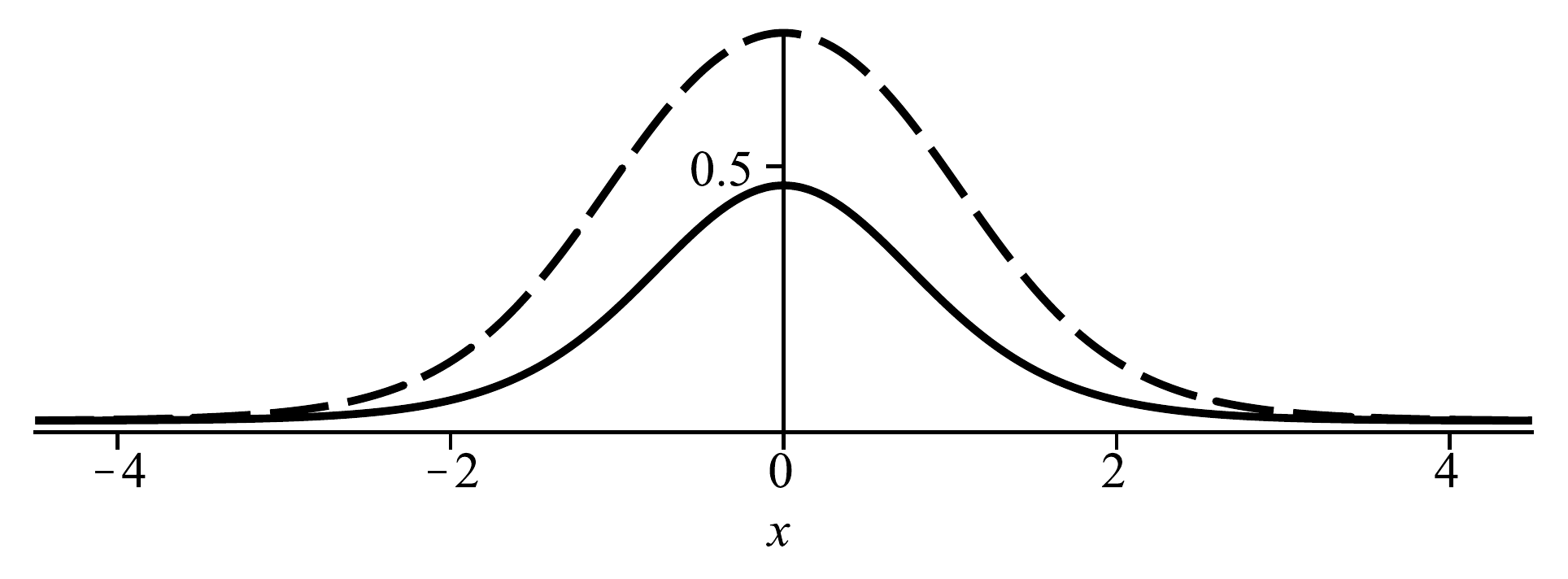}}
\caption{The lump-like solution \eqref{sols}, depicted for $s=0.5$ (solid line) and for $s=1.0$ (dashed line).}\label{fig8}
\end{figure}
\begin{figure}[h!]
\centerline{\includegraphics[{height=3.6cm,width=7cm}]{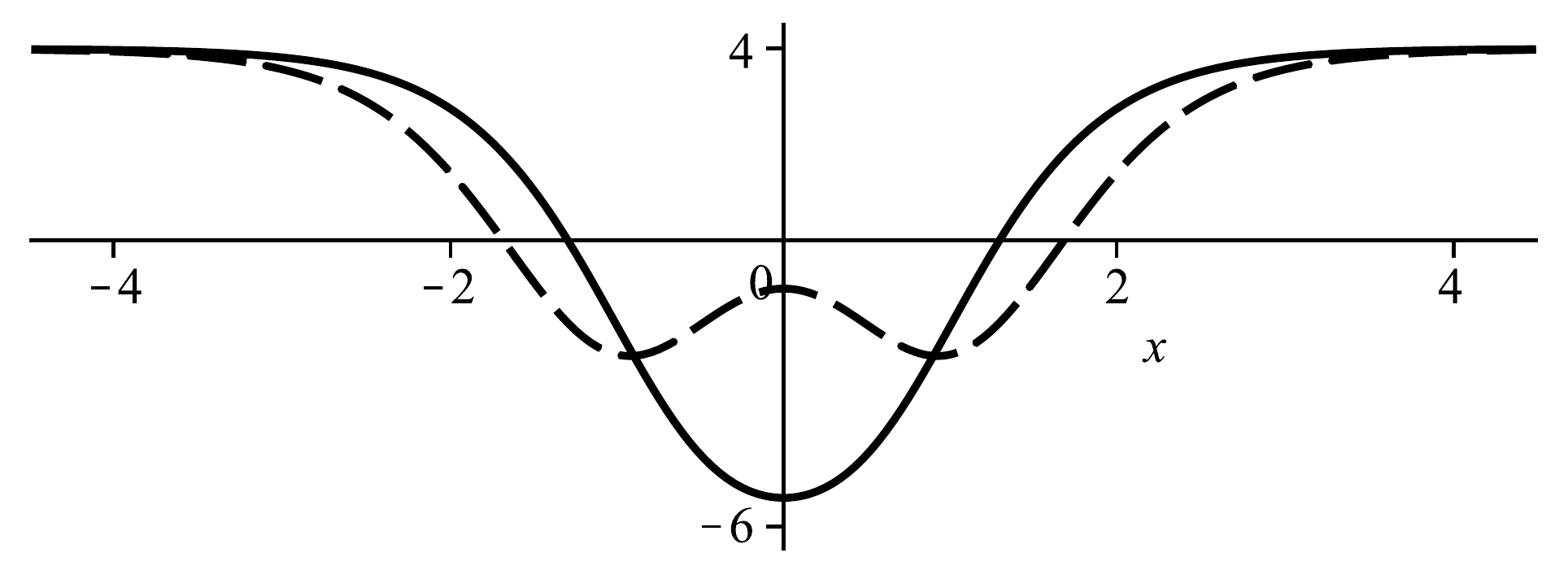}}
\caption{The stability potential \eqref{us}, depicted for $s=0.5$ (solid line) and for $s=1.0$ (dashed line).}\label{fig9}
\end{figure}

Let us now consider the model
\be\label{vs} 
V_s(\phi)=2\phi^2\bigl(\phi-\tanh(s)\bigr)\bigl(\phi-\coth(s)\bigr),
\ee
where $s$ is a nonnegative real parameter. This potential is depicted in Fig.~\ref{fig7}, for some values of $s$. Besides the minimum at $0$, it has another minimum, the global minimum at 
\be  
\phi_+=\frac{3+3\,e^{4s}+\sqrt{e^{8s}+34\,e^{4s}+1}}{4(e^{4s}-1)}\,,
\ee
and a local maximum at
\be
\phi_-=\frac{3+3\,e^{4s}-\sqrt{e^{8s}+34\,e^{4s}+1}}{4(e^{4s}-1)}\,.
\ee  
Although the minimum at $\phi=0$ does not depend on $s$, both $\phi_{\pm}$ depend explicitly on $s$, which also controls the profile of the lump-like solutions. The model was investigated in \cite{ave1} and supports the lump-like solution
\be \label{sols} 
\phi_s(x)=\frac12\bigl(\tanh(x+s)-\tanh(x-s)\bigr),
\ee
which is depicted in Fig.~\ref{fig8}. The parameter $s$ controls the amplitude and width of the lump-like solution, so it is of interest to the current study to investigate its stability. We do this in the standard manner, to get the stability potential
\be\label{us}
u_s(x)=4+24\,\phi_s^2(x)-12\big(\tanh(s)+\coth(s)\big)\,\phi_s(x)\,,
\ee
with $\phi_s(x)$ given by \eqref{sols}.

This is an interesting potential since the value
\be 
\bar{s}=\frac12\ln\left(2+\sqrt{3}\,\right)
\ee 
identifies two regions, one for $s\leq{\bar s}$, where the potential has the single well profile, and the other for $s>\bar s,$ where it develops the shape of a double-well potential, as one illustrates in Fig.~\ref{fig9} for $s=0.5$ and for $s=1.0$. As one knows, the zero mode has the form
\be  
\eta_1(x)=N_1\bigl(\textrm{sech}^2(x+s)-\textrm{sech}^2(x-s)\bigr),
\ee
where $N_1$ is the normalization factor. The zero mode has a node at $x=0$, so there is another bound state with negative energy, which makes the lump-like solution linearly unstable. In general, if the field configuration is known analytically, we can also know the zero mode analytically, since it is the derivative of the field configuration which represents the localized structure, being it topological or non-topological. However, we do not have a general argument to infer if the other bound states could also be expressed analytically.

As one knows, the negative energy bound state is symmetric and can be found as suggested in Ref.~\cite{T1}, for instance, where a nonperturbative procedure to the double-well potential was developed. In several aspects, the potential with $s>{\bar s}$ gives rise to a quantum system that is similar to systems used before to describe the ammonia molecule $NH_3$, thought to be shaped as a symmetrical pyramid, with the Nitrogen tunneling through the planar equilateral triangle of Hydrogen atoms; see, e.g., Refs.~\cite{A1,A2} for two distinct double-well models, and Ref.~\cite{E1} for experimental results. In the model above, one has a single parameter $s$ which controls the gap between the ground state, symmetric, and the zero mode, antisymmetric. So we can use this in applications of current interest in quantum physics. 

\begin{figure}[t]
\centerline{\includegraphics[{height=3.2cm,width=7cm}]{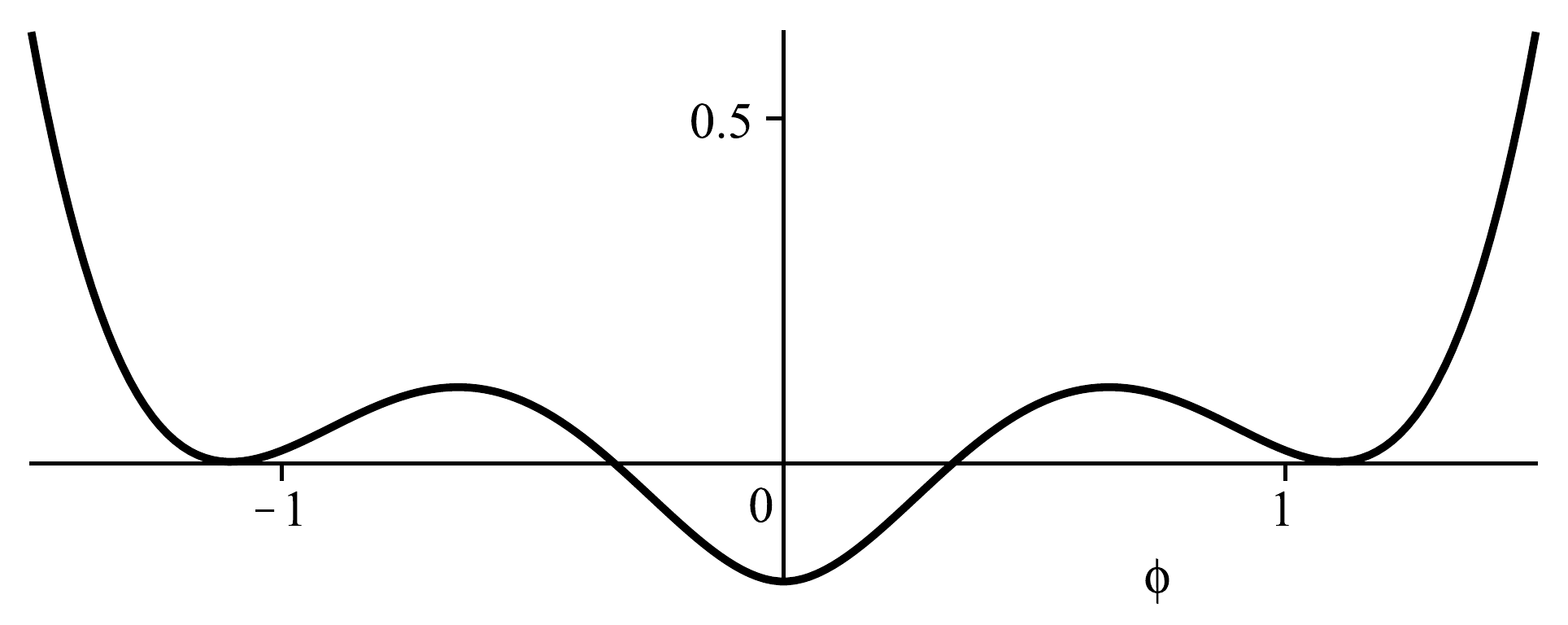}}
\caption{The potential that appears using $a=\phi_+$, depicted for $s=1$.}\label{fig10}
\end{figure}

\begin{figure}[t]
\centerline{\includegraphics[{height=3.4cm,width=7cm}]{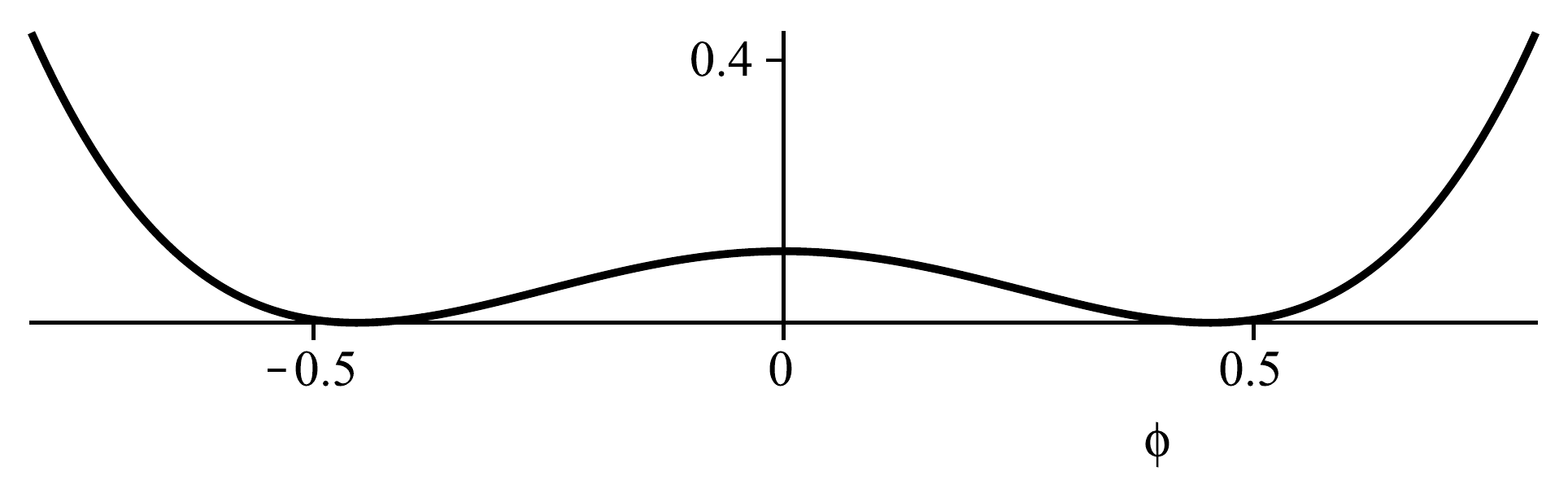}}
\caption{The potential \eqref{Vsa}, that appears using $a=\phi_-$, depicted for $s=1$.}\label{fig11}
\end{figure}

\begin{figure}[t!]
\centerline{\includegraphics[{height=3.8cm,width=7cm}]{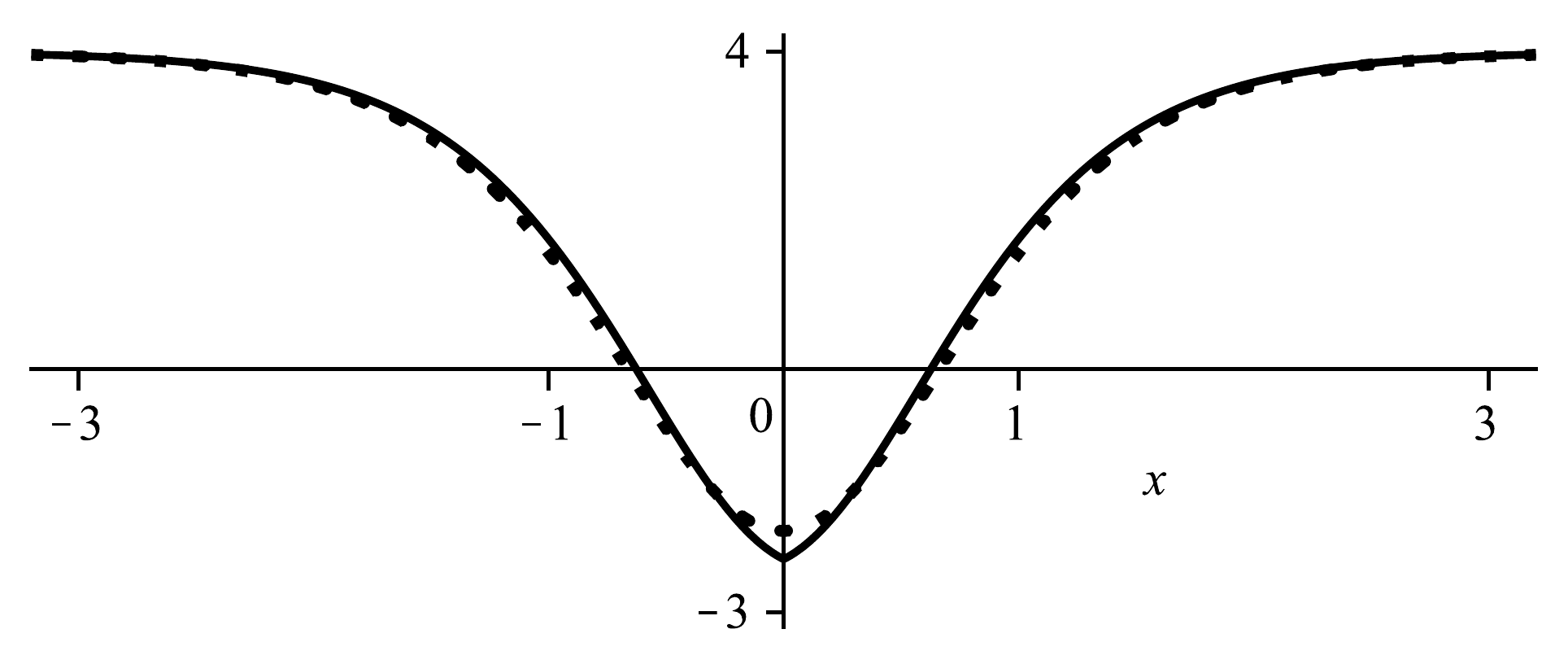}}
\caption{The stability potential that appears from the model \eqref{Vsa}, depicted for $s=1$ (solid line) and the modified P\"oschl-Teller potential (dotted line).}\label{fig12}
\end{figure}

We go on and deform the model with 
\be 
f_a(\phi)=a-|\phi|
\ee
with $a$ being the minimum at $\phi_+$, and this leads us to a model with two identical nontopological sectors, as illustrated in Fig.~\ref{fig10} for $s=1$. But we can also use $a$ as the maximum at $\phi_-$, and this gives rise to the model
\ben\label{Vsa}
U(\phi)&=&2(\phi_{-}-|\phi|)^2\bigl(\phi_{-}-\tanh(s)-|\phi|\bigr)\nonumber\\
&&\times\bigl(\phi_{-}-\coth(s)-|\phi|\bigr),
\een 
which is illustrated in Fig.~\ref{fig11} for $s=1$. It has a single topological sector with kinklike solution
\be
{\phi}(x)=
\left\{
\begin{array}{c}
-\phi_{-}+\phi_s(x-x_0),\;\;\;\;\;x\leq 0\,,\\
\,\,\\
\,\,\,\,\,\phi_{-}-\phi_s(x+x_0),\;\;\;\;\;x\geq 0\,,
\end{array}
\right.
\end{equation}
with $\phi_s(x)$ given by \eqref{sols}, and $x_0$ such that $\phi_s(x_0)=\phi_{-}$.
The stability potential in this case gives rise to a new quantum system, as illustrated in Fig.~\ref{fig12} for $s=1$, with the zero mode given by
\be
{\eta}_0(x)\!=\!
\left\{
\begin{array}{c}
\!\!{\rm sech}^2(x-x_{0}+s)-{\rm sech}^2(x-x_{0}-s),\;x\leq 0\,,\\
\,\,\\
\!
\!\!{\rm sech}^2(x+x_{0}-s)-{\rm sech}^2(x+x_{0}+s),\;x\geq 0\,,
\end{array}
\right.
\end{equation}
which has no node and so is the ground state of the quantum potential, as it usually happens with topological solutions. One notices that the stability potential depicted in Fig.~\ref{fig12} (solid line) is very similar to the modified P\"oschl-Teller potential
$u(x)=4-6 \,{\rm\sech}^2(x)$ which is also depicted in Fig.~\ref{fig12} (dotted line) and known to have two bound states, the zero mode and another one, with positive energy. This suggests that the new potential (Fig.~\ref{fig12}, solid line) has another bound state, with energy close to the value found for the modified P\"oschl-Teller potential. It is always possible to find the bound state numerically and we can, for instance, use the standard shooting method, as recently explored in Ref.~\cite{bbg}. Although the subject of \cite{bbg} was the scattering of kinks of the sinh-deformed $\varphi^4$ model, the model also engenders stability potential similar to the ones depicted in Fig.~\ref{fig12}, which presents a positive bound state with energy close to the energy of the positive bound state which appears in the modified P\"oschl-Teller potential.

We can deform these models again and again, to generate new field theory models, in a way similar to the investigations done before for the other families of models. As shown in \cite{BL} and in the above investigations, the procedure used to deform a given model and construct another one can be implemented algorithmically, to generate distinct models, which can give rise to a replication of the same stability potential, or to generate new potentials, that are not present in the model used to implement the deformation procedure.

\section{Ending comments}
\label{end}

In this work we considered models described by a single real scalar field in $(1,1)$ spacetime dimensions to generate stability potentials that may be used to define new quantum problems. We first used the inverted $\phi^4$ model to show its lump-like structures, which are unstable against small fluctuations. We then deformed the model to generate families of models, having the very same nontopological sectors, but including new topological sectors, all of them having the zero mode and no other bound states.

We also proposed a new family of models that support different lump-like structures that are also unstable against small fluctuations, but we deformed them to get new possibilities, including the presence of kinklike configurations that support a tower of non-negative bound states. The tower of bound states give rise to new quantum systems, whose physical properties will be explored elsewhere. We also found models that can be used to describe quantum potentials with the double-well shape, and so can be used in applications of current interest in quantum mechanics. 

The results of the current work open a new avenue, which instigates us to implement further investigations, in particular, on the construction of new models that support lump-like solutions, as the ones introduced in \cite{ave1,ave2}, for instance. This is of current interest, since lump-like structures may appear in optical fibers as bright solitons \cite{OS1,OS2,OS3}, so we may use the results of the work to describe new bright soliton configurations in fibers. We are also considering the new QM potentials that appeared in Eq.~\eqref{udn}, which are parametrized by $n$ and support $n$ bound states, to see how they behave under the rules of supersymmetric quantum mechanics \cite{Khare,B3,B4}. An investigation of current interest concerns the use of these models to investigate the possibility to construct ladder operators that allow us to connect the several bound states. We believe that this issue will require the presence of shape invariance \cite{B3,B4}, so we are now searching for the possibility to consider the procedure used in this work to describe systems that engender shape invariance. Another interesting issue concerns the scattering of kinks in this model. As one knows, the presence of the zero mode and another positive energy bound state in the modified P\"oschl-Teller potential, makes the scattering of kinks an interesting subject, and we think that the scattering of kinks in these new models may be much more interesting yet, since the presence of extra bound states may induce new effects, as the suppression of two-bounce windows described recently in \cite{argomes}. Another line of investigation concerns issues dealing with the connection between the current procedure and the reconstruction of scalar field theories studied before in \cite{Bor, Vacha,BB2} and in references therein. We are also investigating the possibility to extend the above results to planar structures like vortices and skyrmions. We hope to report on these and other related issues in the near future.\\

\centerline{\bf Acknowledgments}

DB and LL would like to thank Departamento de F\'\i sica Te\'orica, Universidad de Valencia, for the kind hospitality during a short visit that directly contributed to finish this paper.
The work is partially supported by CNPq (Brazil),  by the projects FIS2014-57387-C3-1-P (MINECO/FEDER, EU) and  SEJI/2017/042 (Generalitat Valenciana), the Consolider Program CPANPHY-1205388, and the Severo Ochoa grant SEV-2014-0398 (Spain). DB acknowledges support from project CNPq:306614/2014-6 and LL acknowledges support from project CNPq:307111/2013-0 and CNPq:447643/2014-2. GJO is supported by the Ramon y Cajal project RYC-2013-13019.


\end{document}